\documentclass[aps,prl,reprint,floatfix,superscriptaddress,showpacs]{revtex4-1}
\usepackage[latin1]{inputenc}

\usepackage{color}
\usepackage{amsmath}
\usepackage{amsfonts}
\usepackage{amssymb}
\usepackage{graphicx}
\usepackage{natbib}
\usepackage{multirow}
\usepackage{longtable}
\usepackage{xfrac}
\usepackage{float}
\usepackage[colorlinks=true,linktoc=page,linkcolor=blue,citecolor=blue,filecolor
=blue,urlcolor=blue]{hyperref}
\usepackage[all]{hypcap}
\usepackage[]{units}
\usepackage[normalem]{ulem}

\usepackage{soul,xcolor,colortbl}    
\sethlcolor{green}      

\begin{document}
\title{A stringent upper limit on the direct 3$\alpha$ decay of the Hoyle State in $^{12}$C}
\date{\today}

\author{R. Smith}
\affiliation{Faculty of Science, Technology and Arts, Sheffield Hallam University, Sheffield, S1 1WB, UK}
\affiliation{The LNS at Avery Point, University of Connecticut, Groton, CT 06340-6097, USA}

\author{M. Gai}
\affiliation{The LNS at Avery Point, University of Connecticut, Groton, CT 06340-6097, USA}

\author{M. W. Ahmed}
\affiliation{Triangle Universities Nuclear Laboratory and Department of Physics, Duke University, Durham, North Carolina 27708-0308, USA}
\affiliation{Department of Mathematics and Physics, North Carolina Central University, \\
Durham, North Carolina 27707, USA}

\author{M. Freer}
\affiliation{School of Physics and Astronomy, University of Birmingham, Edgbaston, B15 2TT, UK}

\author{H. O. U. Fynbo}
\affiliation{Department of Physics and Astronomy, Aarhus University, DK-8000 Aarhus C, Denmark}

\author{D. Schweitzer}
\affiliation{The LNS at Avery Point, University of Connecticut, Groton, CT 06340-6097, USA}

\author{S. R. Stern}
\affiliation{The LNS at Avery Point, University of Connecticut, Groton, CT 06340-6097, USA}

\begin{abstract}
\noindent We investigate an implication of the most recent observation of a second $J^\pi = 2^+$ state in $^{12}$C, which was measured using the $^{12}$C($\gamma$,$\alpha$)$^8$Be$_{\textrm{(g.s.)}}$ reaction. In addition to the dissociation of $^{12}$C to an $\alpha$-particle and $^8$Be in its ground state, a small fraction of events (2\%) were identified as direct decays and decays to excited states in $^8$Be. This allowed a limit on the direct 3$\alpha$ partial decay width to be determined as $\Gamma_{3\alpha}~<~32(4)$~keV. Since this 2$^+$ state is predicted by all theoretical models to be a collective excitation of the Hoyle state, the 3$\alpha$ partial width of the Hoyle state is calculable from the ratio of 3$\alpha$ decay penetrabilities of the Hoyle and 2$^+$ states. This was calculated using the semi-classical WKB approach and we deduce a stringent upper limit for the direct decay branching ratio of the Hoyle state of ${\Gamma_{3\alpha} \over \Gamma} < 5.7 \times 10^{-6}$, over an order of magnitude lower than previously reported. This result places the direct measurement of this rare decay mode beyond current experimental capabilities.


\pacs{
21.60.Gx (Cluster models),
26.20.Fj (Stellar helium burning),
21.10.-k (Properties of nuclei; nuclear energy levels),
27.20.+n (6$\leq$A$\leq$19)
}

\end{abstract}

\maketitle




\noindent The Hoyle state at 7.6542 MeV in $^{12}$C is a subject of continued great interest due to the crucial role it plays in the formation of $^{12}$C in the universe, and also due to its peculiar $\alpha$-particle structure \cite{Physics}. Indeed, its very existence was used in ambitious \textit{nuclear lattice simulations} \cite{LatticeEFT} to predict the quark masses and the strength of the electromagnetic interaction in the frame of the anthropic principle.

The discovery and analysis of the broad second $2^+$ state in $^{12}$C \cite{Osaka,iThemba,Yale,Freer,PRL} at approximately 10  MeV has led to an interesting conclusion on the structure of the narrow [$\Gamma$~=~8.5(1)~eV] Hoyle state. This $2^+$ state is predicted to be a collective excitation of the Hoyle state. This is also supported by their similar measured reduced $\alpha$-particle widths and the observation of a Hoyle state rotational band \cite{PRL2,HoyleBand}. The implication of the same underlying structure allows us to study the excited $2^+$ state and subsequently infer properties of the Hoyle state itself.

The Hoyle state is thought to exist as a three $\alpha$-particle system, which \textit{the algebraic cluster model} (ACM) predicts are arranged on an equilateral triangle \cite{ACM,PRL2}. In contrast, \emph{ab initio nuclear lattice simulations} using chiral effective field theory \cite{EFT} predict the three $\alpha$-particles to be arranged on an obtuse triangle. Furthermore, antisymmetrized molecular dynamics (AMD) and fermionic molecular dynamics (FMD) calculations predict a triangular $^8$Be + $\alpha$ configuration \cite{AMDHoyle,FMDHoyle}. It has also been suggested that the Hoyle state could be the nuclear analogue of atomic Bose Einstein Condensation \cite{THSR,BoseEinsteinCondensate} if the 3$\alpha$ system is sufficiently diffuse, allowing the bosonic nature of the $\alpha$-particle to dominate. Each of these models, despite vast differences in their formulations and predicted Hoyle state structures, successfully calculate various experimental observables. Therefore, distinguishing between these models remains a challenge.

Recently, there has been a growing interest in measuring the \emph{direct} 3$\alpha$-decay width of the Hoyle state \cite{Man12,Kir12,ItohHoyle,SmithHoyle,NapoliHoyle,RanaHoyle} in order to learn more about its structure. An upper limit on the direct decay branching ratio of 0.019\% was obtained in the most recent study \cite{RanaHoyle}. 


%
%


For a structureless $\alpha$-condensate, the relative phase spaces and Coulomb decay barriers for direct and sequential decays should entirely determine the branching ratio for the direct 3$\alpha$ decay \cite{KokalovaCondensate}. Under this assumption, a branching ratio of 0.06\% was calculated \cite{SmithHoyle}, larger than the current experimental upper limits. Similar calculations \cite{ZhengHoyleDecay} recently predicted a direct decay branching ratio of 0.0036\%, considerably below current experimental limits. Furthermore, a detailed theoretical analysis into the decay of the Hoyle state, calculated in the Faddeev three-body formalism \cite{Ishikawa}, concluded that the direct decay should contribute at a level $<$ 1\%. The measured experimental enhancement of the $^8$Be$_{\textrm{g.s.}}$ channel relative to some theoretical predictions has been interpreted as an underlying $^8$Be + $\alpha$ structure for the Hoyle state \cite{SmithHoyle}. However, due to theoretical uncertainties, a meaningful comparison with previous data is difficult, since the measured upper limits still lie close to these predictions.

In this paper, we report on new independent analyses of the data reported in reference \cite{PRL}, which were measured at the HI$\gamma$S facility with an optical readout Time Projection Chamber (O-TPC) \cite{JINST}. We have reanalyzed the excitation curve measured for the second 2$^+$ state at approximately 10 MeV and have extracted the decay width of this 2$^+$ state to channels other than $\alpha_0$ + $^8$Be$_{\textrm{g.s.}}$, using the branching ratio published in reference \cite{PRL}. Since the 3$\alpha$ reduced widths of this state and the Hoyle state should be the same, based on the shared underlying structure, their partial widths are related simply by a ratio of decay penetrabilities. We utilise three-body WKB tunnelling calculations to obtain the penetrability ratio and extract an upper limit on the direct 3$\alpha$ decay width of the Hoyle state. The result is compared with predictions of this quantity in the framework of an $\alpha$-condensate structure and we conclude that our experimental upper limit is considerably lower than that expected for an $\alpha$-condensate.

\begin{figure}[h] 
\begin{center}
\includegraphics[width=0.45\textwidth]{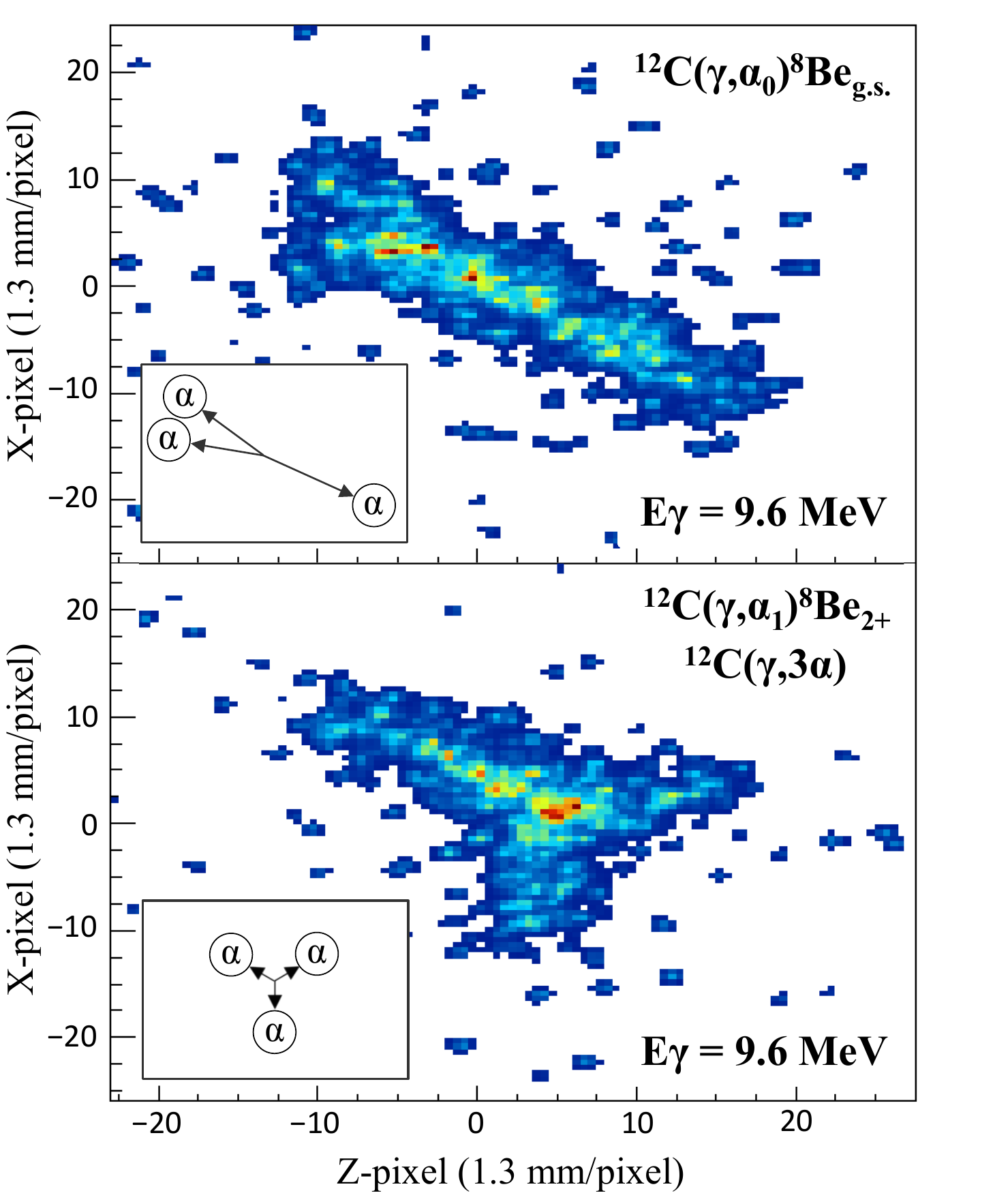}
\end{center}
\vspace{-0.4cm}
\caption{(Color online) Different types of events photographed by the O-TPC for E$\gamma$ = 9.6 MeV. Upper panel: A $^{12}$C($\gamma$,$\alpha_0$)$^8$Be$_{\textrm{(g.s.)}}$ sequential decay. Lower panel: A $^{12}$C($\gamma$,$\alpha_1$)$^8$Be$_{2^+}$ sequential decay or $^{12}$C($\gamma$,3$\alpha$) direct decay, which are indistinguishable.}
\label{fig:OTPCevents}
\end{figure}

In the experimental measurement of the $^{12}$C$(\gamma,3\alpha)$ reaction \cite{PRL} the three outgoing $\alpha$-particles were observed using an O-TPC detector \cite{JINST}, which ``takes a picture" of the event and allows us to visualize it in three dimensions. A reanalysis of the O-TPC data has allowed us to observe two distinctively different $^{12}$C dissociation events: In one class of events, two $\alpha$-particles emerge close together, correlated in space from the decay of the ground state of $^8$Be. These are denoted as $^{12}$C$(\gamma,\alpha_0)^8$Be$_{\textrm{g.s.}}$, and a typical event is shown in the upper panel of Fig. \ref{fig:OTPCevents}. In the second class of events, we observe large opening angles between the three $\alpha$-particles. A typical event is shown in the lower panel of Fig. \ref{fig:OTPCevents}. This means that the $\alpha$-particles share the breakup energy more evenly, and cannot correspond to decays through the $^8$Be$_{\textrm{g.s.}}$. These events compose 2.0(2)\% of the total ``on resonance" events at $E\gamma$~=~10.05 MeV, as previously reported \cite{PRL}.

These rare ``direct decay" events have previously been identified \cite{JINST} as the combined sequential $^{12}$C$(\gamma,\alpha_1)$$^8$Be$_{2^+}$ and direct $^{12}$C$(\gamma,3\alpha)$ dissociation of the 10 MeV second $2^+$ state of $^{12}$C. The two have strongly overlapping kinematics because the $^8$Be$_{2^+}$ is so broad [$\Gamma \approx$ 1.5 MeV]. As highlighted in a recent theoretical study \cite{Refsgaard2018}, decays through the broad $^8$Be$_{\textrm{g.s.}}$ \emph{ghost anomaly} are also indistinguishable from direct 3$\alpha$ decays. Therefore, it can be concluded that


\begin{align}
\Gamma -  \Gamma_{\alpha 0} &= \Gamma_{\alpha1} + \Gamma_{3\alpha},\\
\Gamma_{3\alpha} &< \Gamma -  \Gamma_{\alpha 0}.
\label{eq:Widths}
\end{align}

\noindent where $\Gamma$ is the total width of the second 2$^+$ state, $\Gamma_{\alpha 0}$ and $\Gamma_{\alpha1}$ are the $\alpha$-widths for decays to the $^8$Be$_{\textrm{g.s.}}$ and $^8$Be$_{2^+}$, respectively, and $\Gamma_{3\alpha}$ is the width for direct 3$\alpha$ decay.

The 2$^+$ resonance curve measured using the $^{12}$C($\gamma$,$\alpha_0$)$^8$Be$_{\textrm{(g.s.)}}$ reaction \cite{PRL} has since been reanalyzed using a single level $R$ matrix fit. New resonance parameters were extracted and published in reference \cite{FreerRevModPhys}: $E_R$~=~10.025(50) MeV, $\Gamma~=~1.60(13)$ MeV, $\theta_{\alpha_0}^2$~=~2.3(3) and $\Gamma_\gamma~=~220^{+26}_{-36}$ keV. In the new $R$~matrix analysis, the resonance parameters were found to be sensitive to the channel radius. The penetration factor is evaluated at the channel radius, which is given as $R=R_0(A_1^{1/3} +A_2^{1/3})$, where $A_1$ and $A_2$ are the mass numbers of the $\alpha$ and $^8$Be decay fragments. Therefore, the chosen best fit and corresponding uncertainties include consideration of both the obtained $\chi^2$ values and the systematic variation due to the $R_0$ parameter.

A $\chi^2$ per degree of freedom close to unity was obtained for a large $R_0$ = 1.6 fm. The fit requiring such an unusually large channel radius is strongly indicative of a spatially extended structure. A large $R_0$ is also required to obtain the experimental width of the Hoyle state \cite{BarkerTreacy}. A comparably large $\theta_{\alpha_0}^2$ value is obtained using $R_0$~=~1.6~fm \cite{FreerRevModPhys}. Due to their shared underlying structure, the Hoyle state and 2$^+$ excitation have similar $\theta_{\alpha_0}^2$ values, each requiring a large channel radius.




Using the measured total width of the 2$^+$ state of $\Gamma$~=~1.60(13)~MeV \cite{FreerRevModPhys} and the 2\% observed branching ratio for non-$\alpha_0$ events, we deduce $\Gamma_{3\alpha}~<~32(4)$~keV. From this we extract an upper limit of $\Gamma_{3\alpha}$ for the Hoyle state. In $R$~matrix theory, the partial width of a state, for channel, $i$, is given by $\Gamma_{i} = 2 P_i \gamma_i^2$, where $P_i$ is the penetrability factor and $\gamma_i^2$ is the reduced channel width. Recall that the reduced-channel-width-to-Wigner-limit ratio is denoted by $\theta_i^2$, which is defined as

\begin{equation}
\gamma_i ^2 = \theta_i ^2 \times {3 \over 2} {\hbar ^2 \over MR^2},
\label{eq:ReducedWidths}
\end{equation}




Since the Hoyle state and second 2$^+$ state share the same intrinsic structure, we assume that $\theta_{3\alpha}^2(\textrm{Hoyle}) = \theta_{3\alpha}^2(2^+)$, hence


\begin{align}
\gamma_{3\alpha}^2(\textrm{Hoyle}) &= \gamma_{3\alpha}^2(2^+)\\
\Gamma_{3\alpha}(\textrm{Hoyle}) &= \Gamma_{3\alpha}(2^+)\frac{P_{3\alpha}(\textrm{Hoyle})}{P_{3\alpha}(2^+)}.
\label{eq:WidthComparison}
\end{align}

\noindent We discuss below our method for calculating the ratio of the two penetrability factors in equation \ref{eq:WidthComparison} and we deduce an upper limit on the direct 3$\alpha$ width for the Hoyle state, $\Gamma_{3\alpha}(\textrm{Hoyle})$.

The three $\alpha$-particle system is fully described by their three relative coordinates. Using these relative coordinates, the 3$\alpha$ Coulomb and Centrifugal decay barriers may be parameterised by the hyperradius of the system, $\rho$, as defined by equation 3 in reference \cite{Cocoyoc2019}. It was assumed that only the lowest value partial wave for each decay contributes, since it has the lowest barrier \cite{WKBSchematic,NielsenHypermomentum}.

\begin{figure}[h] 
\begin{center}
\includegraphics[width=0.45\textwidth]{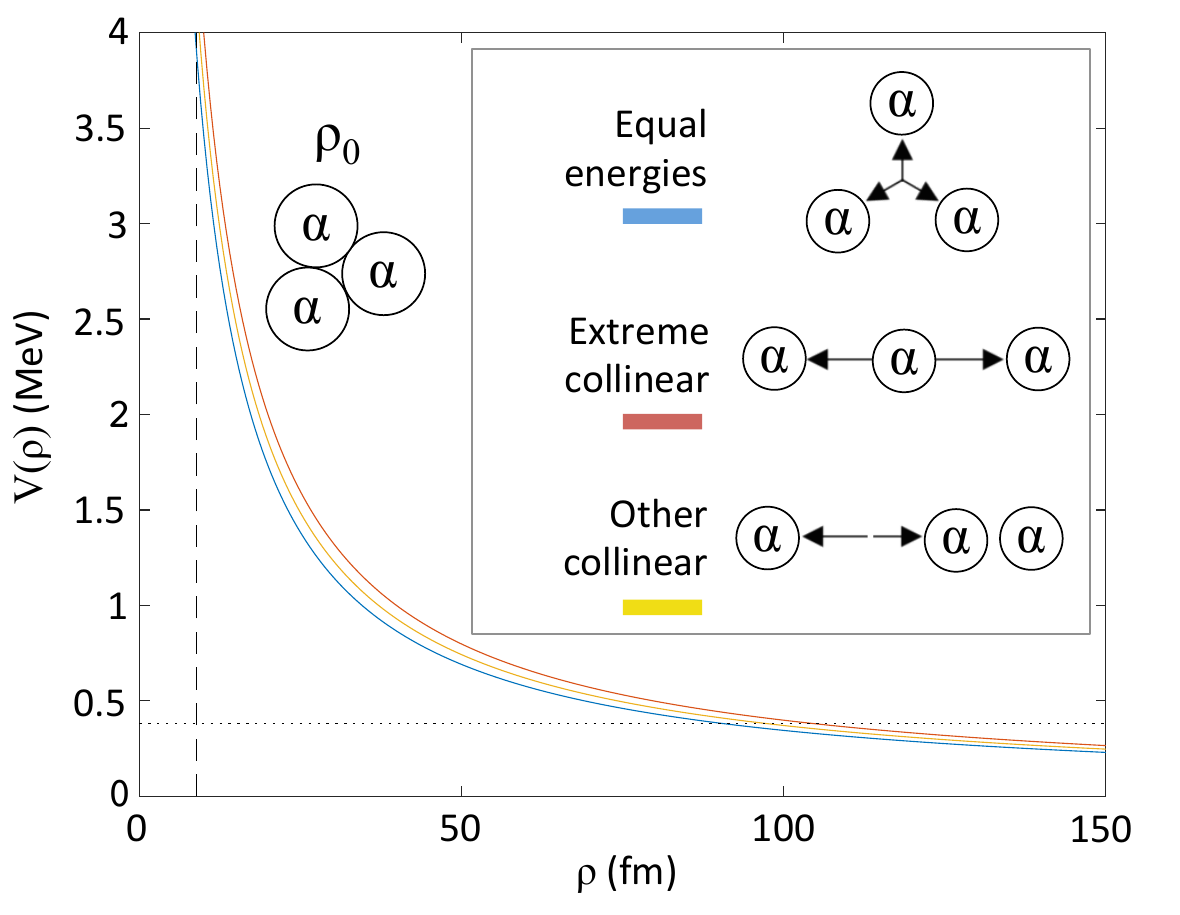}
\end{center}
\vspace{-0.4cm}
\caption{(Color online) The Coulomb potential as a function of the hyperradius for various direct 3$\alpha$ decay configurations of the Hoyle state. The horizontal and vertical dashed lines show the energy above the 3$\alpha$ threshold and the $\rho_0$ value, respectively.}
\label{fig:CoulombBarriers}
\end{figure}

Unlike in a simple sequential decay, the direct decay into three $\alpha$-particles permits a variety of different $\alpha$-configurations. These vary from an extreme collinear decay, where two of the $\alpha$-particles are emitted back-to-back, leaving the third stationary, through to an equal energies case, where the $\alpha$-particles are emitted at 120$^\circ$ to one another. These cases are pictured in the inset of figure \ref{fig:CoulombBarriers}. Importantly, these different decay configurations possess different decay barriers which are shown as a function of the hyperradius, $\rho$, for the Hoyle state in figure \ref{fig:CoulombBarriers}. This must be considered when calculating the penetrability factor for the 3$\alpha$ decay.

The Penetrability of Three Alphas (\texttt{PeTA}) code \cite{PeTA} was used to calculate the 3$\alpha$ penetrability factors for the direct decay of the Hoyle state and 2$^+$ excitation. Further details may be found in reference \cite{Cocoyoc2019}. This Monte-Carlo code uniformly samples the available phase space for a given direct decay and generates 3$\alpha$ momenta subject to the constraints of energy and momentum conservation. The code calculates the penetrability factor for the 3$\alpha$ decay using the semi-classical WKB approach presented in reference \cite{WKBSchematic}, and an average is found over all of the generated 3$\alpha$ configurations. This approach assumes that the trajectories of the three $\alpha$-particles and the barrier do not change during tunneling. Full consideration of the decay dynamics and the 3$\alpha$ Coulomb interactions will perturb this static picture.

\begin{figure}[h] 
\begin{center}
\includegraphics[width=0.5\textwidth]{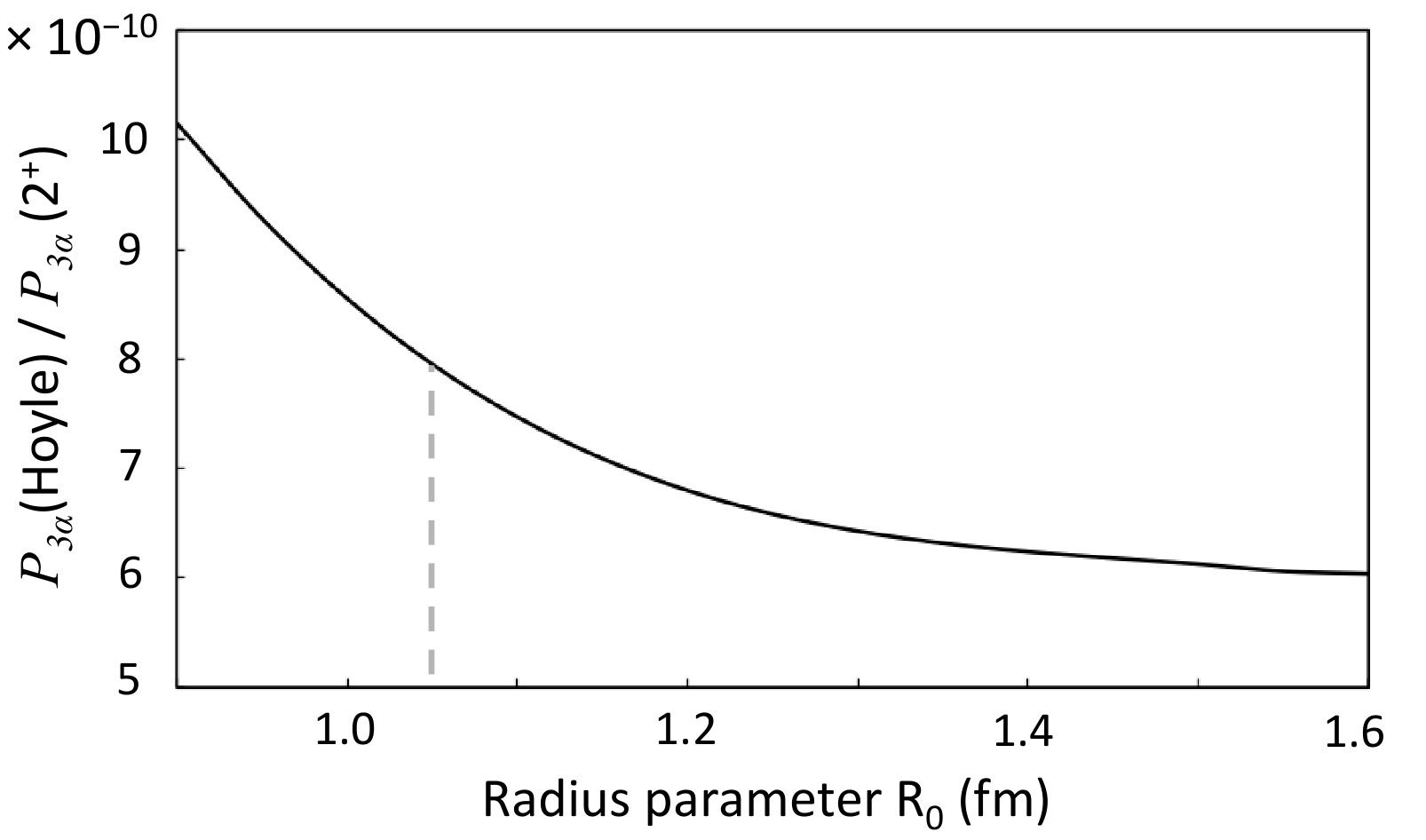}
\end{center}
\vspace{-0.4cm}
\caption{The ratio of the 3$\alpha$ penetrability factors as a function of the $R_0$ parameter. The vertical dashed line marks the $R_0$ value that best reproduces the experimental $\alpha$-particle radius.}
\label{fig:PenetrabilityRatios}
\end{figure}

The \texttt{PeTA} WKB code \cite{PeTA} has previously been used to predict the 3$\alpha$ phase space distribution for decays from the Hoyle state \cite{SmithHoyle,Cocoyoc2019}. Its agreement with another model \cite{Refsgaard2018}, which accounts for the 3$\alpha$ Coulomb interactions differently, has been noted. Using this code, the ratio of penetrability factors $P_{3\alpha}(\textrm{Hoyle})/P_{3\alpha}(2^+)$ was found to be 8(2)~$\times$~10$^{-10}$. The $\alpha$-particles are calculated to tunnel from a point where they are just touching ($\rho_0$ in figure \ref{fig:CoulombBarriers}) out to the classical turning point. Therefore, the result depends on the $\alpha$-particle radius, which scales with the $R_0$ parameter. Although the penetrabilities vary strongly with $R_0$, their ratio, $P_{3\alpha}(\textrm{Hoyle})/P_{3\alpha}(2^+)$, remains fairly insensitive to this parameter, as shown in figure \ref{fig:PenetrabilityRatios}.

Including a more realistic Coulomb interaction for overlapping spherical charge distributions, rather than point charges, will affect the result. However, this has previously been evaluated for the sequential decay of the Hoyle state and was found to give a modest change for reasonable $R_0$ values \cite{EFB24Proc}. Furthermore, the inclusion of an attractive nuclear potential will act to lower the tunneling barrier near to the channel radius, thus altering the absolute values of the decay penetrability. However, we already account for a variation in barrier height by analyzing the change in $P_{3\alpha}(\textrm{Hoyle})/P_{3\alpha}(2^+)$ over a range of $R_0$. Therefore, the approximate effect of an attractive potential is included within this uncertainty.

Using equation \ref{eq:WidthComparison}, and the ratio of penetrabilities, an upper limit on the 3$\alpha$ partial width was calculated to be 2.6(7)~$\times$~10$^{-5}$~eV. Using this value and the known width for the Hoyle state, $\Gamma$~=~8.5(1)~eV, we deduce a direct decay branching ratio upper limit of $B.R. < 5.7 \times 10^{-6}$. All experimental uncertainties and systematic errors from the calculations have been included in this limit, totalling around 25\%. The primary assumption of this work, that the reduced 3$\alpha$ channel widths of the Hoyle state and 2$^+$ state are the same, adds a further factor of 2 uncertainty. This is based on the $\theta_{\alpha_0}^2$ values of 1.5 \cite{FreerRevModPhys} and 2.3(3) for the Hoyle and second 2$^+$ states, respectively.

In the most recent experiment, an upper limit on this direct decay branching ratio was placed at 0.019\% ($\approx$10$^{-4}$)  \cite{RanaHoyle}. Therefore, the upper limit deduced in this work is approximately two orders of magnitude below the experimental limit. Our calculated upper limit, which is based on the measured properties of the 10 MeV second 2$^+$ state, yield, for the first time, an upper limit for direct decay far lower than existing theoretical predictions \cite{Ishikawa,SmithHoyle,ZhengHoyleDecay}. In references \cite{SmithHoyle,KokalovaCondensate}, it was stated that, in the popular $\alpha$-condensate description of the Hoyle state, the direct decay branching ratio should be around 0.06\%. Similar calculations \cite{ZhengHoyleDecay} recently predicted a direct decay branching ratio of 0.0036\%. The presently derived upper limit is considerably lower than these calculations, and is therefore inconsistent with the predictions for an $\alpha$-condensate state. As also noted in reference \cite{SmithHoyle}, an experimental enhancement of the sequential decay channel could indicate a $^8$Be + $\alpha$ structure for the Hoyle state as predicted by the molecular dynamics calculations \cite{AMDHoyle,FMDHoyle}.

In addition, future measurements of the 3$\alpha$ direct decay of the Hoyle state may also provide new insight into the nature of this 2$^+$ resonance. If direct decay of the Hoyle state is experimentally measured at a level higher than our deduced upper limit, this could imply that the Hoyle and 2$^+$ resonances have different $\theta_{3\alpha}^2$ values, and hence different structures, leading to important theoretical re-evaluations. Another possibility could be the existence of additional 2$^+$ strength at energies higher than the 11.2 MeV maximum measured at HI$\gamma$S, which would have distorted the single level $R$ matrix fit used in this work. A future incompatible result would therefore prompt further experimental work.

\begin{acknowledgments}
The authors would like to thank the staff of HI$\gamma$S at Triangle Universities Nuclear Laboratory for the operation of the facility, and C.~R.~Howell for in-depth discussions and assistance with various aspects of the work. This material is based on work supported by the U.S. Department of Energy, Office of Science, Office of Nuclear Physics grants DE-FG02-94ER40870, DE-SC0005367, DE-FG02-97ER41033 and DE-FG02-91ER-40608, and the UK Science and Technology Facilities Council.
\end{acknowledgments}

\vspace{-0.4cm}



\end{document}